\documentclass [12pt]{article}
\usepackage{amssymb}
\usepackage{amsmath}
\usepackage{microtype}

\usepackage{type1cm}        
\usepackage{makeidx}         
\usepackage{graphicx}        
\usepackage{multicol}        
\usepackage[bottom]{footmisc}

\usepackage{newtxtext}       %
\usepackage{newtxmath}       

\usepackage{wrapfig}

\newcommand{\be}{\begin{equation}}
\newcommand{\ee}{\end{equation}}
\newcommand{\beal}{\begin{aligned}}
\newcommand{\eeal}{\end{aligned}}
\newcommand\bea{\begin{eqnarray}}
\newcommand\eea{\end{eqnarray}}
\newcommand{\bec}{\begin{cases}}
\newcommand{\eec}{\end{cases}}

\def\bi{\begin{itemize}}
\def\ei{\end{itemize}}
\begin{document}


\title{Primordial Black Holes and Higgs Vacuum Decay}

\author{Ruth Gregory}

%
%
\date{}

\maketitle

\center{Department of Physics, King's College London, 
The Strand, London, WC2R 2LS}

\abstract{Phase transitions are part of everyday life, yet are also believed to be part 
of the history of our universe, where the nature of particle interactions change as the 
universe settles into its vacuum state. The discovery of the Higgs 
\cite{ATLAS:2012ae,Chatrchyan:2012tx}, and measurement 
of its mass suggests that our vacuum may not be entirely stable, and that a further 
phase transition could take place. This article is based on a talk in the Oldenberg Series, 
and reviews how we find the probability of these phase transitions, discussing past work 
on how black holes can dramatically change the result! Apart from a brief
update at the end, this article mostly follows the content of the talk.}

\begin{center}

{Contribution to: Springer Lecture Notes in Physics 

\emph{Gravity, Cosmology, and Astrophysics:
A Journey of Exploration and Discovery with Female Pioneers} 

\emph{Ed: Betti Hartmann 
and Jutta Kunz}

}

\end{center}

\newpage

\section{Executive Summary}
\label{sec:exec}

This work was motivated by the observation that phase transitions typically are seeded 
by impurities, yet the techniques used by theorists 
\cite{Kobzarev:1974cp,coleman1977,callan1977,CDL} 
to compute the probability of decay
are extremely idealised - with huge mathematical simplifications assumed in order to
make computations tractable. In the case of the Higgs vacuum, one intriguing 
possibility is that the self coupling of the Higgs could become negative at large
Higgs values \cite{Degrassi:2012ry,Bezrukov:2014ina,Ellis:2015dha,Blum:2015rpa}, 
leading to the conclusion that we live in a metastable vacuum, so
the question of just how accurate these idealised computations are becomes of
very direct relevance! In the work that I review
\cite{GMW,Burda:2015isa,Burda:2015yfa,Burda:2016mou}, we take the simplest possible
impurity for seeding vacuum decay: a black hole. This breaks the symmetry of
the standard theoretical description, yet maintains sufficient theoretical control that
the computations can be done largely analytically. The punchline of this article is that 
black holes change the computation....\emph{enormously}!

\section{The Coleman Computation}
\label{sec:intro}

In this 
introductory section, I motivate and review the Euclidean method used to compute 
tunnelling amplitudes in field theory. This discussion is based largely on the series
of papers by Coleman \cite{coleman1977,callan1977,CDL}

\subsection{Motivation of the Euclidean method}

The phenomenon of tunnelling is a uniquely quantum mechanical one -- classically, 
if a particle does not have enough energy to scale a barrier, it will remain forever on one side; 
quantum mechanically however there is always a small probability of it being found 
across this seemingly unscalable hurdle. Calculating the probability of this process is one
of the first computations we meet in Introductory Quantum Mechanics -- we study the
time independent Schr\"odinger equation in one spatial dimension
\be
E \Psi = - \frac{\hbar^2}{2m} \frac{d^2 \Psi}{dx^2} + V(x) \Psi 
\ee
\begin{figure}
\begin{center}
\includegraphics[scale=.6]{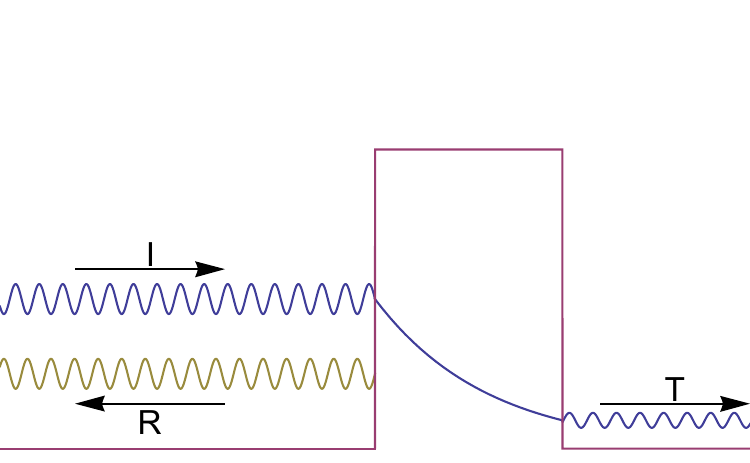}~~~~
\caption{A sketch of the 1+1-dimensional Schr\"odinger tunnelling calculation.}
\label{fig:schr}       
\end{center}
\end{figure}
for a potential that has a simple, square, barrier:
\be
V(x) = \bec
0 & x<0 \;\;\; \& \;\;\; x>d\\
V_0 & 0<x<d
\eec
\ee
where $V_0>E$.
The solution for $\Psi$ is oscillatory outside the barrier, and exponential underneath it
(see figure \ref{fig:schr}).
\be
\Psi(x) = \bec
I e^{i\omega x} + R e^{-i \omega x} & x<0 \\
A e^{\Omega x} + B e^{-\Omega x} & 0<x<d \\
T e^{i\omega x}  & x>d
\eec
\ee
where $\omega^2 = 2m E/\hbar^2$, and $\Omega^2 = 2m(V_0-E)/\hbar^2$.
Continuity of $\Psi$ and it's derivative at each side of the barrier gives 4 boundary
conditions, allowing the wave function to be solved completely. The probability that
a particle will tunnel through the barrier is therefore
\be
\frac{|T|^2}{|I|^2} = \left [ 1 + \frac{V_0^2 \sinh^2 \Omega d}{4 E(V_0-E)}\right]^{-1}
\sim e^{-2\Omega d}
\ee
which is strongly dominated by an exponential factor representing the strength of
the barrier the particle has to tunnel through:
\be
\Omega d = \frac1\hbar \int_0^d \sqrt{2m(V_0-E)} dx
\ee
Computing this leading order exponential suppression is the aim of the Euclidean method.

Now consider the following problem. A classical particle is at the tip of a square well of
depth $\Delta V$, it falls in, transferring the potential to kinetic energy, transits the well and 
goes up the other side. Using $\frac12 m \dot{x}^2 = \Delta V$, we have
\be
\int \sqrt{2m\Delta V} dx = 
\int \sqrt{2m\Delta V} \dot{x} d\tau = 
\int 2\Delta V d\tau = 
\int \left ( \Delta V + \frac12 m \dot{x}^2\right) d\tau = S_E
\ee
\begin{figure}[t]
\begin{center}
\includegraphics[scale=.5]{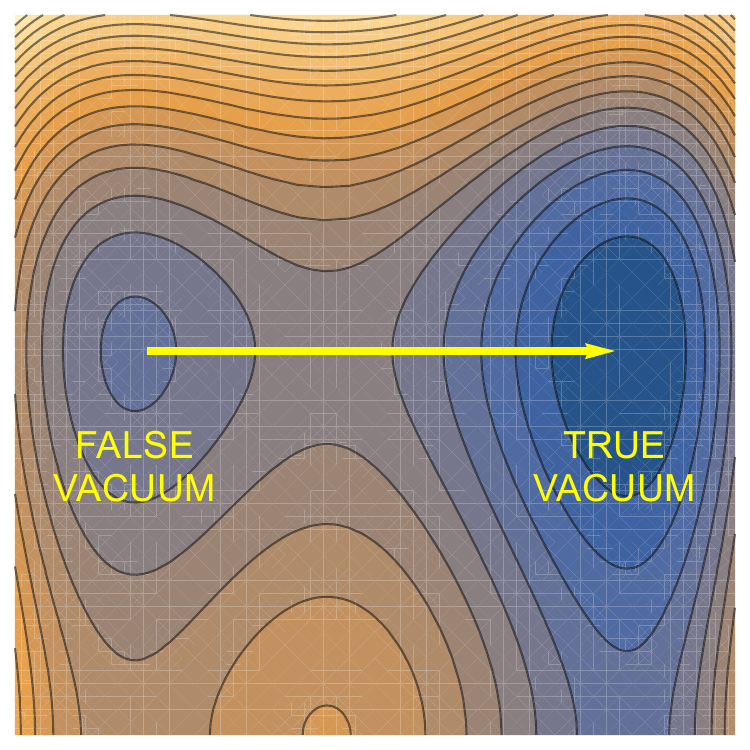}~~~~
\caption{The ``escape'' path from the local false vacuum is a saddle point in 
the inverted potential space.}
\label{fig:bbw}       
\end{center}
\end{figure}
but this is just the Euclidean action for the motion of said point particle!
For more general potentials, this gives an intuitive visualisation of the 
tunnelling amplitude calculation. We invert the potential and consider
the classical motion of a ``particle'' from the (now unstable) local maximum
to an exit point, and back again. This motion was called the \emph{bounce} 
\cite{coleman1977}, and the one-dimensional tunnelling can readily be extended
to multiple dimensions, with the bounce becoming a saddle trajectory between the local 
maximum (of the inverted Euclidean potential) to an exit point. This \emph{most 
probable escape path} picture, developed by Banks, Bender, and Wu \cite{Banks:1973ps} 
(see figure \ref{fig:bbw})
led naturally to the Euclidean field theory approach of Coleman and others
\cite{Kobzarev:1974cp,coleman1977}.

\subsection{Tunnelling in field theory}

In Quantum Field Theory (QFT) fields typically have an interaction potential, with
the minimum of the potential representing the vacuum of the theory. It might be, however,
that this minimum is not unique, either due to nonlinearly realised symmetry, or because
there is more than one minimum in the potential separated in field space.
Since the vacua are at distinct values of the field, the process of tunnelling from one 
minimum to another with lower overall energy will give rise to a \emph{first order} phase
transition, which we expect to proceed via bubble nucleation. 

\begin{figure}[t]
\begin{center}
\includegraphics[scale=.5]{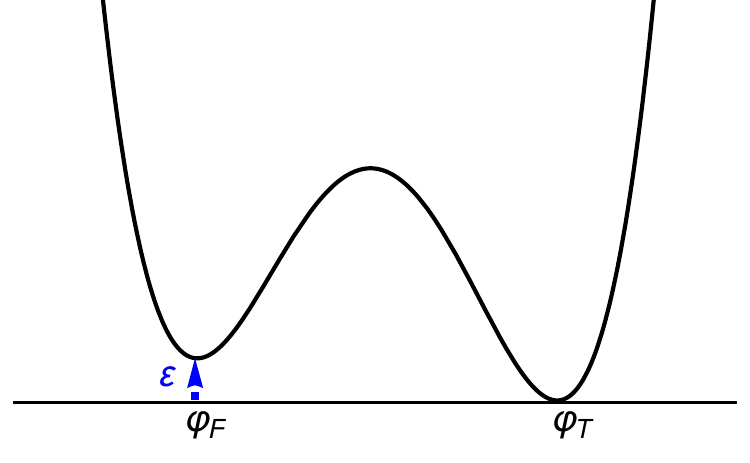}
\caption{A sketch of the type of potential relevant for the instanton computation.}
\label{fig:TF}       
\end{center}
\end{figure}
To find the probability of nucleation, we can follow the Euclidean prescription, 
taking the equations of motion and analytically continuing to Euclidean time,
$t\to i\tau$, finding a solution known as an \emph{instanton} which describes the 
formation of a bubble of true vacuum inside the false. For simplicity,
I will discuss this process for a simple scalar field
\be
{\cal L}_\varphi = \frac12 (\partial\varphi)^2 - V(\varphi)
\ee
where $\varphi$ is a real scalar field, and $V(\varphi)$ is a potential with
a false (local minimum) and true (global minimum) vacuum (see figure \ref{fig:TF}),
where the difference in energy ($\varepsilon$) is assumed small 
relative to the potential barrier to be traversed. As is conventional, I have set
$c=\hbar=1$ in the ensuing discussion.

Before outlining the full procedure, it is instructive to consider the physically motivated 
simplification of the problem: the \emph{Goldilocks Bubble}. Consider a bubble of 
radius $R$ that fluctuates into existence. The bubble will cost energy to form, as
between the vacua the field has to transit from $\varphi_F$ to $\varphi_T$, passing
through a region with high potential and gradient energy; this gives rise to a \emph{wall}
which will carry energy momentum, hence ``cost'' to form. On the other hand, the 
interior volume has dropped from energy $\varepsilon$ to zero energy. If the bubble
is small, the surface area to volume ratio means the gain from energy will be too small to 
sustain the bubble and it will recollapse. If the bubble is large, then the energy budget 
will allow the bubble to grow, but it will be ``expensive'' to form in the first place -- there
is therefore a ``just right'' bubble size where the bubble is initially at rest at formation.
Putting the details in, and writing $\sigma$ for the energy per unit area of the bubble wall
we see that the energy budget is
\be
\delta E = 2\pi^2 R^3 \sigma - \frac{\pi^2}{2} \varepsilon R^4
\ee
recalling that we are in 4 Euclidean dimensions so that the bubble is a 3-sphere.
$\delta E$ is stationary at $R = 3\sigma/\varepsilon$, for which
$\delta E = 27 \pi^2 \sigma^4/2\varepsilon^3$, hence we expect that the
bubble will nucleate at this radius with a probability
\be
{\cal P} \sim e^{-27 \pi^2 \sigma^4/2\varepsilon^3\hbar}
\ee
It turns out that this \emph{thin wall} intuition of Coleman is extremely efficient
at extracting the essence of the full computation.

To outline the full computation, one must solve the Euclidean field equations
\be
\frac{1}{\rho^3} \frac{d~}{d\rho} \left [ \rho^3 \frac {d\varphi}{d\rho} \right]
+ \frac{\partial V}{\partial \varphi} =0
\ee
where $\rho^2 = \tau^2 + {\bf x}^2$ is the distance from the origin in Euclidean 
$\mathbb{R}^4$.
Taking the potential to have the approximate form
\be
V(\varphi) \approx \frac\lambda4 (\varphi^2 - \eta^2)^2 - \frac{\varepsilon}{2\eta} 
(\varphi - \eta)
\ee
where $\varepsilon \ll \lambda \eta^4$, then $V$ has two local minima, approximately 
at $\pm \eta$,
with the vacua energies separated by $\varepsilon$. To leading order, the equation 
of motion for $\varphi$ is
\be
\varphi'' + \frac{3 \varphi'}{\rho} + \lambda \varphi (\varphi^2-\eta^2) = 0
\ee
which can be solved numerically with the boundary conditions $\varphi \to -\eta$
as $\rho \to \infty$, $\varphi(0) \approx \eta$. Performing this integration, the solution
is well approximated by
\be
\varphi \simeq \eta \tanh \left [  \sqrt{\lambda/2}\, \eta (R-\rho) \right]
\label{approxwall}
\ee
where $R\approx 3\sigma/\varepsilon$, and the energy per unit area of the wall is given by
\be
\beal
\sigma &= \frac{1}{R^3} \int_0^\infty \rho^3 \left [ \frac12 \varphi^{\prime2} 
+ V(\varphi)\right] d\rho= \frac{\lambda \eta^4}{2R^3} \int_0^\infty \rho^3 \text{sech}^4 \left [ 
\sqrt{\frac\lambda2}\, \eta (R-\rho) \right] d\rho\\
&\approx \sqrt{\frac\lambda2} \frac{4\eta^3}{3}
\eeal
\ee
This analytic approximation is excellent for $\sqrt{\lambda/2}\, \eta R \gg 1$, and even 
for potentials that have far thicker bubble walls it gives a very good ballpark estimate
for the tunnelling probability.

\subsection{Tunnelling with Gravity}

In the previous discussion, a key feature was that the vacuum energy was different 
for the true and false vacua, but we know that energy gravitates, hence the false vacuum 
will have a different gravitational behaviour than the true vacuum, and the bubble wall
will also have a gravitational signature. The impact of gravity was first worked out
in the paper of Coleman and de Luccia \cite{CDL} using the thin wall approximation
discussed above. This is based on the seminal work of Israel \cite{Israel}, describing
the gravitational effects of thin shells (in this case, the bubble wall).

First, it is worth making a few remarks about the Euclidean approach with gravity.
Below the Planck scale, we expect that spacetime is essentially classical, but that 
gravity can contribute to quantum effects through the impact of spacetime curvature,
and the back-reaction of quantum fields on the spacetime. Usually, we take some
fixed classical background and quantise around this, an approach used in black
hole thermodynamics and cosmological perturbation theory. There is a broader sense
in which we can use gravity semi-quantum mechanically however, and that is by
using the partition function approach of Gibbons and Hawking \cite{gibbonshawking}. 
While the philosophy of
this requires some finessing, the basic methodology is clear: we extend the
partition function to include the Einstein-Hilbert action, 
\be
S_{GH} = - \frac1{16\pi G} \int_{\cal M} d^4 x \sqrt{g} (R-2 \Lambda) 
+ \int d^4 x {\cal L}_{\text{matter}}
\left ( + \frac1{8\pi G} \int_{\partial{\cal M}} d^3 x \sqrt{h} K \right)
\label{GibHawk}
\ee
where the Gibbons-Hawking boundary term has been included (in brackets) for completeness.
At finite temperature we have a finite periodicity of Euclidean time, and typically
we integrate over geometries with the same boundary conditions at $\partial {\cal M}$. 
Whereas there is no clear method for dealing
with fluctuations, the saddle points of the path integral are unambiguous -- these are
solutions of the classical Euclidean field equations.

\begin{figure}[t]
\begin{center}
\includegraphics[scale=.45]{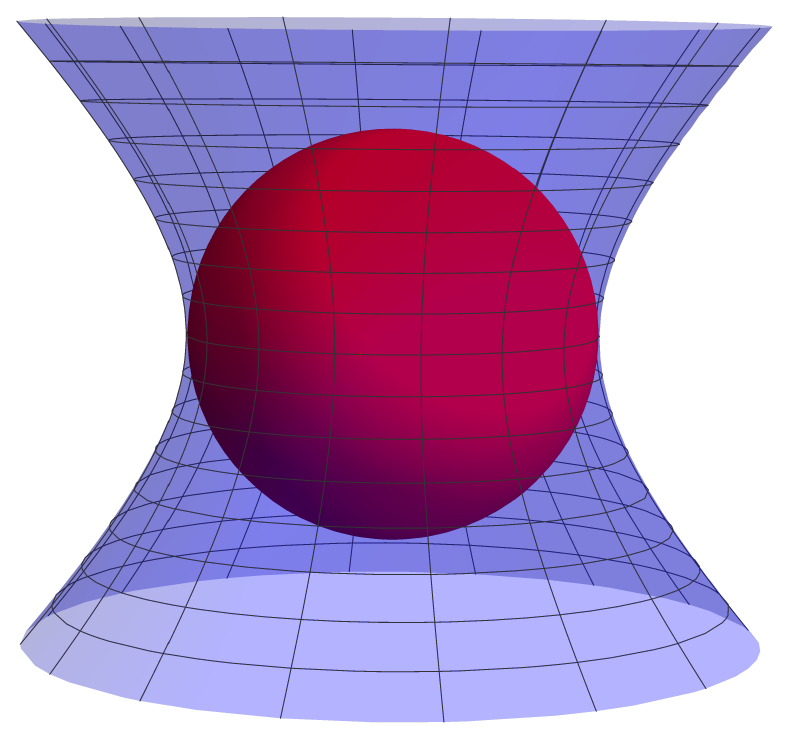}\hskip 1cm
\includegraphics[scale=.34]{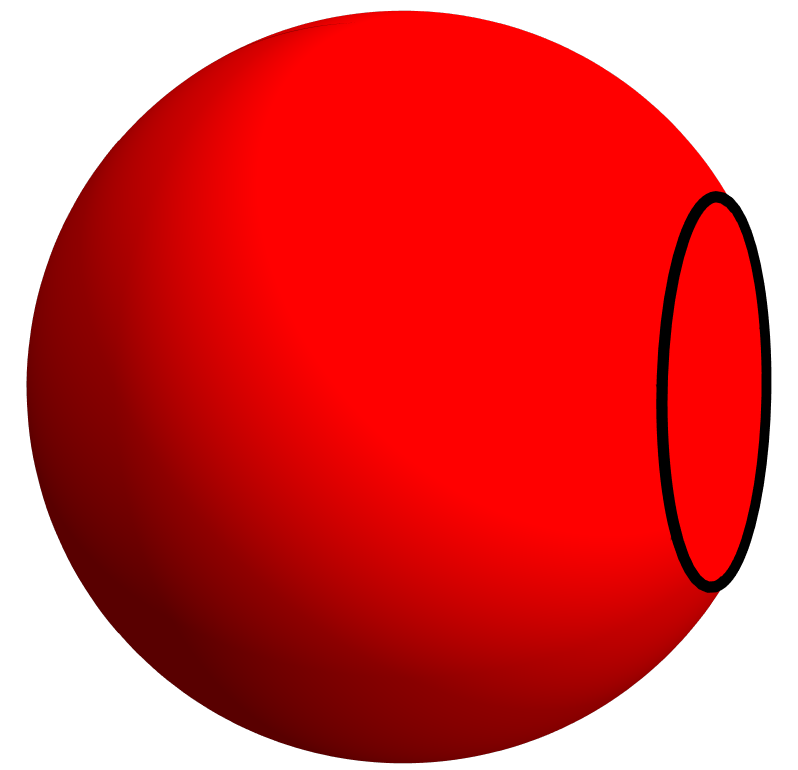}
\caption{{\bf Left}: Lorentzian and Euclidean de Sitter space.
{\bf Right}: the instanton replaces the cap of the sphere with a flat surface. 
The kink at the interface represents the energy momentum of the bubble wall.}
\label{fig:cdl}      
\end{center}
\end{figure}

First I will give a qualitative picture of the instanton before describing the CDL 
calculation in more detail. For pictorial simplicity, consider the case of tunnelling 
from a finite vacuum energy $\varepsilon$ to zero vacuum energy. 
A finite positive vacuum energy is a positive cosmological constant, which we 
know to be de Sitter spacetime. This has a Lorentzian description in terms of the 
surface of a hyperboloid embedded in a five-dimensional spacetime:
\be
w^2 + x^2+y^2+z^2-t^2 = \ell^2 = \frac{3}{8\pi G\varepsilon}
\ee
here, $\ell$ is the scale of curvature of the de Sitter spacetime, and is related 
as shown to the false vacuum energy.
On rotation to Euclidean time, this becomes a 4-sphere
embedded in $\mathbb{R}^5$ (see figure \ref{fig:cdl}). Zero vacuum energy on 
the other hand corresponds to Minkowski spacetime, and is just flat, or planar.
Our instanton must therefore cut a cap off the de Sitter 4-sphere and replace it with
a flat surface as illustrated in figure \ref{fig:cdl}.

We can play the same ``Goldilocks bubble'' game as before, but noting from 
\eqref{GibHawk} that there is potentially a gravitational contribution to energy. 
First, the cost of the wall, coming from the matter contribution in \eqref{GibHawk}, 
is the same: ${\cal E}_{\text{wall}} = 2 \pi^2 R^3 \sigma$. 
Next, the naive energy gain from false to true vacuum depends on the volume of 
the cap excised from the sphere, and this is captured by the bulk contribution 
to \eqref{GibHawk}:
\be
{\cal E}_{\text{cap}}(R) = 
\frac23 \pi^2 \varepsilon \ell^4 \left [ 2 - 2\left ( 1 - \frac{R^2}{\ell^2} \right)^{\frac32}
+ 3 \frac{R^2}{\ell^2}
\right] 
\ee
Using just these two terms gives an approximate answer that includes the 
curved geometry of de Sitter space, but it neglects the impact of the (negative) 
gravitational potential energy due to the gravitational field of the wall. To capture
this, we need to look more closely at the extrinsic curvature of the wall, as this
effect is understood as the contribution from the Gibbons Hawking boundary term
in the action.

In the Coleman de Luccia (CDL) approach \cite{CDL}, they describe the instanton in the 
thin wall limit, using the Israel conditions \cite{Israel}. The wall separates the false 
vacuum, de Sitter space:
\be
ds^2 = d\rho^2 +  \ell^2 \sin^2 (\rho/\ell) d\Omega_{I\!I\!I}^2
\ee
from the true, Minkowski, vacuum
\be
ds^2 = dr^2 + r^2 d\Omega_{I\!I\!I}^2 
\ee
The Israel equations relate the jump in the extrinsic curvature across the wall to the
energy in the wall. The wall sits at $r_0 =\ell \sin(\rho_0/\ell) = R$, with normal 
$d\rho$ / $dr$. The wall is characterised by an energy-momentum tensor that is 
proportional to the induced metric on the wall,
\be
h_{ab} = g_{ab} - n_a n_b
\ee
and the wall extrinsic curvature is given by:
\be
K_{ab} = - \Gamma^c_{ab} n_c = 
\bec
\frac{\cos(\rho_0/\ell)}{\ell \sin(\rho_0/\ell)} & \text{FV}\\
\frac{1}{r_0} & \text{TV}
\eec
\ee
Substituting into the Israel equations relevant for the Euclidean wall gives
\be
K_{\alpha\beta}^+ - K_{\alpha\beta}^- = -\frac1R \left ( 1 - \sqrt{1 - \frac{R^2}{\ell^2} }
\right) h_{\alpha \beta} = - 4 \pi G \sigma h_{\alpha\beta}
\ee
Writing $\bar{\sigma} = 2\pi G\sigma$ for compactness of notation, this is easily seen
to be solved by
\be
R_0 = \frac{4\bar{\sigma}\ell^2}{1 + 4 \bar{\sigma}^2 \ell^2}
\label{RCDL}
\ee
We can then substitute this solution into the full Euclidean action
\be
\beal
{\cal B} &= \frac{1}{16\pi G} \int_{\text{cap}} (R-2\Lambda) \sqrt{g} d^4 x 
+ \int_{\text{wall}}\left [ \frac{ (K_+-K_-)}{8\pi G} +\sigma \right ] \sqrt{h} d^3x\\
&= {\cal E}_{\text{cap}}(R) +  {\cal E}_{\text{wall}}(R) + {\cal E}_{\text{grav}}(R) 
\eeal
\ee
where we have written the integrals for general $R$ to make connection to the 
Goldilocks argument, and identify the gravitational potential energy of the wall curvature 
for general $R$ as:
\be
{\cal E}_{\text{grav}}(R) = \int_{\text{wall}} \frac{ (K_+-K_-)}{8\pi G} \sqrt{h} d^3x=
- \frac{2 \pi R^2}{4 G} \left [ 1 - \left ( 1 - \frac{R^2}{\ell^2} \right)^{\frac12}
\right] 
\ee
Making this action stationary with respect to $R$ recovers the Coleman de Luccia $R_0$
as expected, and gives the tunnelling exponent
\be
{\cal B}_{CDL}(R_0) = \frac{\pi \ell^2}{G} \frac{16 \bar{\sigma}^4 \ell^4}
{(1 + 4 \bar{\sigma}^2 \ell^2)^2}
\label{BCDL}
\ee

Why is this computation important for the Higgs vacuum?
In particle physics, we describe fundamental interactions via the Standard Model (SM)
Lagrangian, which encodes the bosons, fermions, and their interactions.
\be
\beal
{\cal L}_{SM} = &-\frac12 \text{Tr} \, {\bf G}_{\mu\nu} {\bf G}^{\mu\nu}
- \frac12 \text{Tr}\,  {\bf W}_{\mu\nu} {\bf W}^{\mu\nu} - \frac14 F_{\mu\nu}F^{\mu\nu}\\
&+ \left ( {\bf D}_\mu \Phi \right) ^\dagger {\bf D}^\mu \Phi 
+ \mu^2 \Phi^\dagger \Phi - \frac\lambda2 \left ( \Phi^\dagger \Phi \right)^2
+ ..... 
\eeal
\ee
A key feature of this is the Higgs scalar field and its self coupling, or the Higgs
potential. The self coupling of the Higgs, $\lambda$ acquires radiative corrections,
and changes with energy scale $V(\Phi) \sim \lambda(\Phi) |\Phi|^4/4$. As was realised
some years ago \cite{1982Natur.298,Krive:1976sg,Lindner:1988ww,Sher:1988mj},
this could have implications for the stability of the Higgs vacuum. The calculation
depends on the masses of other fundamental particles, particularly the top quark
(see \cite{CMS:2019esx,ATLAS:2019guf} for more recent results)
and the values of the mass of the Higgs and the top quark put us in a region where
the self-coupling could potentially become negative at very high energies
\cite{Isidori:2001bm,EliasMiro:2011aa,Degrassi:2012ry,Gorsky:2014una,
Bezrukov:2014ina,Ellis:2015dha,Blum:2015rpa}.

Whether or not we should be concerned at the metastability of the Higgs vacuum
then becomes an issue of computing the probability of decay. If we use this 
Coleman de Luccia result, the half life of decay is around $10^{138}$ years, 
well in excess of the age of the universe! We might therefore think this metastability
is not an issue, but this would be to ignore the fact that the calculation I have just
described is incredibly idealised -- very much a ``spherical cow''! It is therefore
time to revisit our understanding of vacuum decay, and to explore whether we
can take a step towards a more realistic set-up, including the effect of an impurity.

\section{Thin Wall Bubbles with Black Holes}
\label{sec:bh}

In this section, I review how including a black hole as a seed for false vacuum
decay changes the above picture, and the key results from 
\cite{GMW,Burda:2015isa,Burda:2015yfa,Burda:2016mou}
(see also \cite{Tetradis:2016vq,Chen:2017suz,Mukaida:2017bgd,Gorbunov:2017fhq}). 
I will stick with the thin wall approximation, as this allows a largely analytic approach, 
and review the numerical work in the next section.

\subsection{Bubbles with black holes}

The Coleman de Luccia bubbles discussed in the last section are incredibly
ideal -- the universe is completely smooth and featureless, akin to a supercooled
fluid. Since we are discussing the gravitational effect of the vacuum, we should also
think about what happens if the universe is not featureless and isotropic. The simplest
possible impurity we can think of is a black hole. This is an exact solution to Einstein's
gravitational equations, and can be added to the picture without adding any further matter
content. We also know how to treat a black hole in Euclidean space -- indeed, this was
one of the most dramatic conceptual discoveries of the Gibbons-Hawking paper 
\cite{gibbonshawking}:
that a Euclidean black hole naturally has a periodic time, and demanding non-singularity
of the Euclidean geometry mandates a periodicity of Euclidean time $\Delta \tau = 8\pi GM$.
Given that the partition function for a field at temperature $T$ is described by a period
(Euclidean) time coordinate with periodicity $\beta = 1/T$, this naturally leads to the 
conclusion that a black hole has temperature $T = 1/8\pi G M$ -- consistent with Hawking's
breakthrough result just two years earlier \cite{Hawking:1975vcx}. Of course, there 
are niggles with this interpretation, not least that a black hole is not in thermal 
equilibrium with its surroundings,
but the correspondence between this Euclidean approach and the considerably more 
complex Lorentzian computation is compelling, and to this day we compute the temperature
of a black hole by looking at its Euclidean continuation. 

Now consider the geometrical set-up. The idea is that we have a black hole in the false
vacuum that acts as a seed for vacuum decay. A bubble will therefore nucleate around
the black hole, leaving the same exterior solution, but replacing the interior with 
the true vacuum and, potentially, a remnant black hole that will in general have a 
different mass to the original seed. We are therefore looking for solutions for a thin
bubble wall that match two different vacuum energies (i.e.\ cosmological constants) 
and two different masses of black hole:
\be
ds^2 = f_\pm(r) d\tau_\pm^2 + \frac{dr^2}{f_\pm(r)} + r^2 d\Omega_{I\!I}^2
\label{bhbub}
\ee
where 
\be
f_\pm (r) = 1 - \frac{2GM_\pm}{r} - \frac{\Lambda_\pm}{3} r^2
\ee
is the metric potential for a black hole with a cosmological constant. One
point of confusion can be around the meaning of $M_+$ in the instanton space,
where the only \emph{black hole} is the one at the centre of the solution, with
horizon determined by $M_-$. However, recall that gravity in four dimensions
propagates -- we experience the impact of the moon's gravity here on Earth via
the tides, and indeed the sun's gravity via the neap and spring differences in tide. 
In other words, we can deduce locally the masses of these celestial objects by
looking at local tidal forces. It is exactly the same with the exterior $M_+$ mass --
we detect the original seed mass because the tidal forces outside the bubble are
still those of the original seed. 

The next remark to make when looking for an exact solution in GR is whether
we are imposing an Ansatz on the geometry, or whether our solution is general.
Mathematically, introducing a black hole alters the symmetry of the
CDL set-up, with $SO(4)$ symmetry, by breaking the equivalence between spatial and 
timelike directions, leaving an $SO(3)\times U(1)$ symmetry: the spherical symmetry of
the black hole we see in space, and the $U(1)$ of the periodic Euclidean time direction.
Fortunately, the problem of the exact solution was solved some time ago in the 
context of braneworld models, where a general Birkhoff theorem was proven for 
braneworlds (a more general case of the thin wall) \cite{BCG}, 
meaning that the solution described above in \eqref{bhbub} is indeed a fully 
general solution to the equations of motion of a wall with the desired symmetry.

\begin{figure}[t]
\begin{center}
\includegraphics[scale=.6]{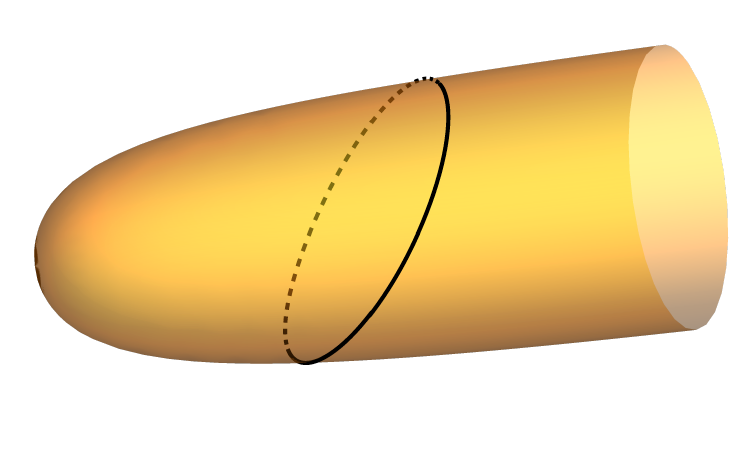}
\caption{A sketch of a wall wrapping the periodic Euclidean time direction. Note that
this is a very special solution where the periodicity of the wall motion precisely matches
that of the black hole.}
\label{fig:wall-bh}       
\end{center}
\end{figure}
Returning to the Goldilocks bubble concept, many of the components are similar, yet 
there is a crucial difference. The general bubble will have a trajectory $R(\tau)$ around
the Euclidean time circle which now must be determined from the Israel equations.
There is also another key difference around background subtraction, described presently,
meaning that there is no alternative but to roll up our sleeves and proceed with the
calculation.

The following discussion paraphrases the analysis in \cite{Burda:2015yfa}, picking
out the key calculational steps.
The bubble is described by a trajectory in both the plus and minus Euclidean 
black hole spacetimes \eqref{bhbub}
\be
X^a = (\tau_\pm(\lambda),R(\lambda),\theta,\phi)
\ee
where we parametrise the trajectory by the local time, $\lambda$ on the bubble wall,
i.e.
\be
f_\pm \dot{\tau}^2_\pm + \frac{\dot{R}^2}{f_\pm} = 1
\label{propertime}
\ee
where dots denote derivatives with respect to $\lambda$.
Taking normals that point towards increasing $r$,
\begin{equation}
n_\pm={\dot \tau}_\pm\,dr_\pm-\dot r\,d\tau_\pm,\label{normal}
\end{equation}
(noting ${\dot\tau}_\pm\geq0$ for orientability), the Israel junction conditions are
\begin{equation}
f_+\dot{\tau}_+-f_-\dot{\tau}_-=-2 \bar{\sigma} R.\label{jnc}
\end{equation}
We can use \eqref{propertime} to rewrite this as a Friedman-like equation 
for $\dot R$:
\begin{equation}
{\dot R}^2 = 1-\left ( {\bar\sigma}^2 + \frac{\bar\Lambda}{3} 
+ \frac{(\Delta\Lambda)^2}{144{\bar\sigma}^2} \right) R^2 
-\frac{2G}{R} \left ( {\bar M} + \frac{\Delta M \Delta\Lambda}
{24{\bar\sigma}^2}\right) - \frac{(G\Delta M)^2}{4 R^4 {\bar\sigma}^2},
\label{rdoteqn}
\end{equation}
where $\Delta M=M_+-M_-$ is the black hole mass difference, and 
$\bar M=(M_++M_-)/2$ the average, with similar expressions for $\Lambda$.
There are also two accompanying equations for the time coordinates on each
side of the wall, 
\be
f_\pm\dot{\tau}_\pm = \frac{\Delta \Lambda}{12\bar{\sigma}}R +
\frac{\Delta M}{2\bar{\sigma} R^2} \mp \bar{\sigma}R
\ee
which can be integrated once $R(\lambda)$ is found.

This gives an effective potential for the motion of the brane that allows us
to quickly characterise the motion, which is easily found numerically. The generic
solution is periodic in $\lambda$, therefore we are now in the interesting position 
of having to fix the periodicity of Euclidean time not by the black hole mass, but
by the wall trajectory.

\subsection{Computing the bounce action}

Having found this wall trajectory, we then need to compute the 
Euclidean action of the thin wall instanton:
\be
\beal
I_E = &-\frac{1}{16\pi G} \int_{{\cal M}_+} \sqrt{g} 
({\cal R}_+-2\Lambda_+)
-\frac{1}{16\pi G} \int_{{\cal M}_-} \sqrt{g} 
({\cal R}_--2\Lambda_-)\\
&+ \frac{1}{8\pi G} \int_{\partial{\cal M}_+} \sqrt{h} K_+ 
- \frac{1}{8\pi G} \int_{\partial{\cal M}_-} \sqrt{h} K_- 
+ \int_{\cal W} \sigma \sqrt{h}
\eeal
\ee
and subtract the action of the background
\be
I_{\text{FV}} = -\frac{1}{16\pi G} \int_{{\cal M}} \sqrt{g} 
({\cal R}_+-2\Lambda_+)
\ee
to find the bounce action for the decay amplitude.

In these expressions, ${\cal M}$ is the geometry of the background false vacuum 
seed space, ${\cal W}$ is the location of the wall in the instanton geometry, which is
a submanifold of ${\cal M}$. ${\cal M_-}$ is the true vacuum part of the instanton 
geometry inside the bubble wall, and ${\cal M_+}$ the part exterior to the bubble, with 
$\partial{\cal M}_\pm$ refers to the boundaries of each of ${\cal M_\pm}$ induced by 
the wall, each of which have a normal pointing outwards to larger $r$, in agreement
with the Israel prescription. In general, as described in \cite{Burda:2015yfa}, there 
may also be additional boundary or bulk terms required for renormalisation of the action.

Having noted these expressions, a very important observation is that in order to perform
the background subtraction of the false vacuum geometry, we must have the \emph{same}
geometries at large $r$ (or outside the wall). This means that ${\cal M}_+$ must be
\emph{identical} to the portion of ${\cal M}$ that lies outside the wall. We can therefore write
\be
\beal
{\cal B} = &\frac{1}{16\pi G} \int_{{\cal M},r<R} \sqrt{g} 
({\cal R}_+-2\Lambda_+)
-\frac{1}{16\pi G} \int_{{\cal M}_-} \sqrt{g} 
({\cal R}_--2\Lambda_-)\\
&+ \frac{1}{8\pi G} \int_{\partial{\cal M}_+} \sqrt{h} K_+ 
- \frac{1}{8\pi G} \int_{\partial{\cal M}_-} \sqrt{h} K_- 
+ \int_{\cal W} \sigma \sqrt{h}
\eeal
\label{fullbounce}
\ee
where hopefully the notation is self explanatory.

Now let us consider the implication of the periodicity of the general solution $R(\lambda)$,
$\beta$, versus the natural periodicity of a black hole spacetime 
\be
\beta_0 = \frac{4\pi}{f(r_h)} = \frac{4\pi r_h}{ 1- \Lambda_\pm r_h^2}
\ee
having used $f(r_h)=0$ at the horizon to simplify the expression. If $\beta\neq\beta_0$,
then we will have a conical singularity at the centre of the Euclidean space $r=r_h$.
In early work on tunnelling with black holes \cite{PhysRevD.35.1161,Berezin:1987ea},
the implication of this mild singularity was neglected, as it was believed to be unimportant,
however it is reasonably straightforward to compute, as explained in Appendix A of
\cite{GMW}. In essence, one smooths out the conical singularity with a family of regular
metrics with curvature confined to proper distance $\rho<\epsilon$ from the `horizon', 
the degree of the singularity in the Ricci scalar is $\rho^{-1}$,
whereas the volume element scales as $\rho$. The nett result is that the integral
\be
\int_{\rho<\epsilon} d^4x {\cal R} \sqrt{g} = 2 {\cal A}\delta 
\ee
is finite, and proportional to the deficit angle $\delta$, and the area of the horizon
${\cal A} = 4\pi r_h^2$. To compute the deficit angle, note that in the natural black
hole geometry, the angular coordinate $\theta_0 = 2\pi \tau/\beta_0$ has a range 
$[0,2\pi]$. With periodicity $\beta$ however, this changes to $[0,2\pi \beta/\beta_0]$,
hence the deficit angle is $\delta = 2\pi - 2\pi \beta/\beta_0$. We therefore see that
the contribution to the action coming purely from the conical deficit is
\be
I_{\text{conical}} = -\frac{1}{16\pi G} \int_{\rho<\epsilon} d^4 x {\cal R} \sqrt{g} 
= - \frac{(\beta_0-\beta)}{\beta_0} \frac{ {\cal A}}{4G}
\ee
Returning to the general expression \eqref{fullbounce}, the first line, which is the
bulk contribution, can now be computed as
\be
\beal
{\cal B}_{\text{bulk}} = &\frac{\Lambda_+}{6G} \int_0^\beta d\tau_+ (R^3-r_{h,+}^3)
-\frac{\Lambda_-}{6G}  \int_0^\beta d\tau_- (R^3-r_{h,-}^3)\\
&+\frac{(\beta_+-\beta)}{\beta_+} \frac{ {\cal A}_+}{4G}
-\frac{(\beta_--\beta)}{\beta_-} \frac{ {\cal A}_-}{4G}\\
=&  \frac{({\cal A}_+- {\cal A}_-)}{4G} 
- \frac{1}{4G} \int d\lambda R^2 \left ( f'_+ \dot{\tau}_+- f'_- \dot{\tau}_- \right) \\
&- \frac\beta{4G} \left [ \frac{ {\cal A}_+}{\beta_+} - \frac{ {\cal A}_-}{\beta_-} - 2 G (M_+ -M_-)
+ \frac{2\Lambda_+}{3} r_{h,+}^3  - \frac{2\Lambda_-}{3} r_{h,-}^3 \right ]
\eeal
\ee
However, noting that $\beta_\pm = 4\pi / f'_\pm(r_{h,\pm})$, and
using the horizon relation $f_\pm(r_{h,\pm})=0$, the term in square brackets on the
final line vanishes.

We are now left with the boundary and wall terms, which can be combined into
an expression dependent on the wall trajectory:
\be
{\cal B}_{\text{wall}} = \int_{\cal W}\hskip -2mm \sqrt{h} \left [ \frac{(K_+ - K_- )}{8\pi G} 
+ \sigma \right ] = -\frac{\sigma}2 \int_{\cal W} \hskip -2mm \sqrt{h}
=\frac{1}{2G} \int d\lambda\,R\left (f_+ {\dot\tau}_+ - f_- {\dot\tau}_-\right)
\ee
having used $f_+ {\dot\tau}_+ - f_- {\dot\tau}_- = -2{\bar\sigma}R$.

Putting all of these results together, the action of the
bounce is:
\begin{equation}
{\cal B}= {{\cal A}_+\over 4G}-{{\cal A}_-\over 4G} +{1\over 4G}
\oint d\lambda \left\{\left ( 2Rf_+ - R^2 f_+'\right)\dot{\tau}_+
- \left (2Rf_- -R^2 f_-'\right) \dot{\tau}_-\right\}
\label{bounceaction}
\end{equation}

\begin{figure}[t]
\begin{center}
\includegraphics[scale=.48]{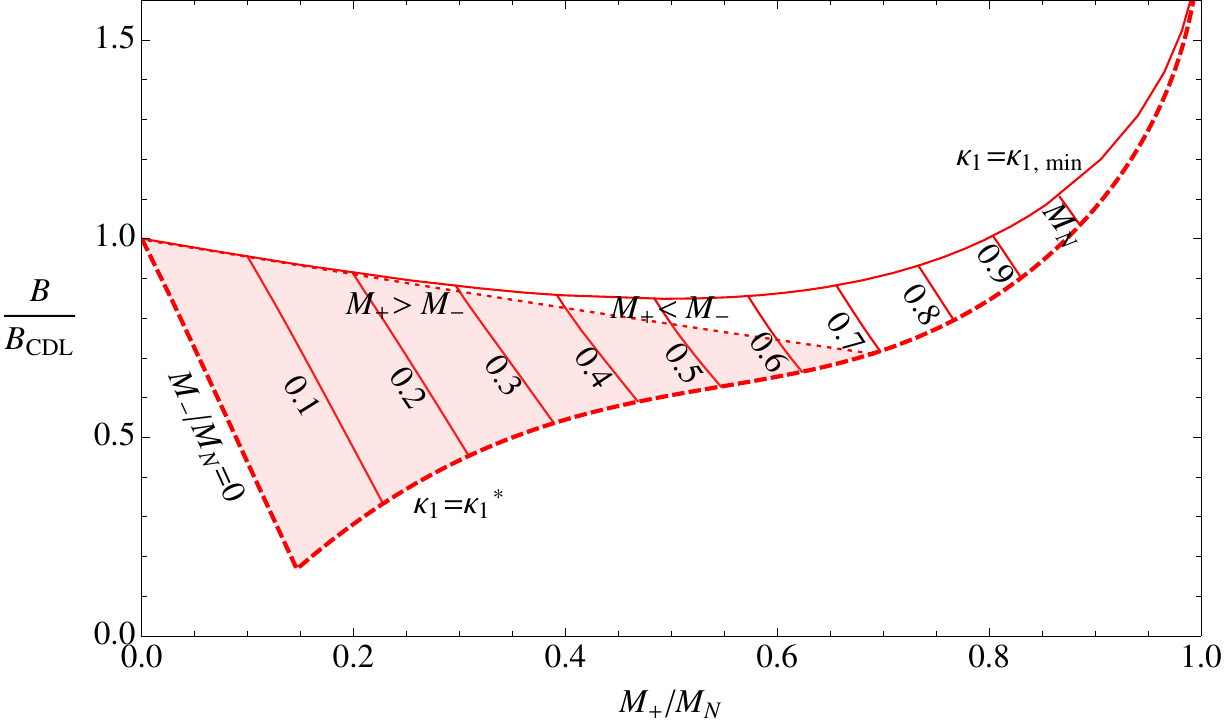}
\caption{A comparison of the CDL bounce action and the black hole seeded action for 
tunnelling from positive to zero vacuum energy. (Fig 5 from \cite{GMW}.)
The seed mass is expressed as a ratio of $M_+$ to the maximum allowed mass in de 
Sitter, the Nariai mass $M_N$, and the values of the remnant mass are labelled. The 
$\kappa$ annotations refer to combinations of the seed and remnant parameters and
are given in \cite{GMW}.}
\label{fig:cfcdl}       
\end{center}
\end{figure}

This computation was the core result of 
\cite{GMW,Burda:2015isa,Burda:2015yfa,Burda:2016mou}. In general, for
a given seed mass, \eqref{rdoteqn} will have bubble solutions for a range of
remnant masses $M_-$. However, evaluating \eqref{bounceaction} shows
that these have different bounce actions, hence different decay rates. For a
given seed mass there will be a unique remnant mass that has the lowest action.
Since the probability of decay is the exponent of the negative of this action,
the decay is dominated by this minimum action value. Figure \ref{fig:cfcdl} 
(Fig.\ 5 from \cite{GMW}) shows how the bounce action depends on the seed
mass for tunnelling from a positive to zero vacuum energy for a value of wall
tension $\bar{\sigma}\ell = 0.2$. For each seed mass (expressed as a ratio
of $M_+$ to the maximum allowed mass in de Sitter, the Nariai mass $M_N$) there is
a range of allowed remnant masses. The minimal action configuration is shown as a 
dashed red line. 

Explicit computation shows that for small seed masses, the minimal action bounce
removes the seed black hole is altogether but it still $\tau-$dependent. For larger 
seed masses, the minimal action bounce is static, leaving behind a remnant black
hole, with increasing remnant mass as the seed mass is increased. These two
branches meet at a critical mass with the lowest action bounce which is both
static, and without a black hole remnant. We can read off from the parameter range
discussion in the appendix of \cite{Burda:2015yfa} the value of $M_+$ at which
this occurs. For example, if $\Lambda_-=0$, we get:
\be
GM_+ = \frac{128\bar{\sigma}^3}{3(\Lambda_++12 \bar{\sigma}^2)^2}
\ee
or,
\be
\frac{M_+}{M_p} = \frac{128\pi}{27} 
\frac{(\bar{\sigma} \ell/2)^3}{(1+4\bar{\sigma}^2 \ell^2)^2}
\frac{\ell}{L_p}
\ee
where $M_p$ and $L_p$ are the Planck mass and length scale respectively.
In order to remain well below the Quantum Gravity scale, we require $M_+/M_p\gg1$,
and we have the bound $\bar{\sigma} \ell<1/2$, hence we require large $\ell/L_p$ with
$\bar{\sigma} \ell$ not too small. However, both  $\bar{\sigma}$ and $\ell$ are 
determined by the properties of the scalar potential, and these conditions are not
particularly realistic, and for the Higgs potential, not relevant at all. Hence, the most
likely bounce is one which has seed mass much larger than this critical value, and
is therefore static and with a remnant black hole.

The most important take home message of this section is that adding a black hole 
decreases the bounce action, hence \emph{enhances} decay. 
The parameters, ($M_\pm$, $\Lambda_\pm$, $\bar\sigma$) used to produce figure 
\ref{fig:cfcdl} are all approximately of a similar order, close to unity, therefore the 
suppression of the bounce action is not as marked as it will be for more realistic 
parameter values. We will see this when exploring the Higgs vacuum in the next section.
However, before discussing the implications
for the Higgs vacuum, I will present two analytic examples of bubbles, first 
re-deriving the CDL result from the perspective of the static patch, which relies
crucially on the computation of the conical singularity integral, then the result
for static instantons.

\subsubsection{CDL revisited}

The CDL instanton has $M_+=M_-=0$, and let $\Lambda_-=0$ so that we are tunnelling
from a positive false vacuum energy $\varepsilon = \Lambda_+/8\pi G = 3/8\pi G \ell^2$ 
to the Minkowski true vacuum as before. With these parameter values, 
\eqref{rdoteqn}, simplifies considerably,
\be
{\dot R}^2 = 1-\left ( {\bar\sigma} + \frac{1}{4{\bar\sigma}\ell^2} \right)^2 R^2 
=1 - \omega^2 R^2
\ee
where $\omega$ is defined above, together with the equations for $\tau_\pm$
\be
\left ( 1 - \frac{R^2}{\ell^2} \right) \dot{\tau}_+ 
= \left ( \omega - 2\bar{\sigma} \right ) R\quad; \qquad
\dot{\tau}_- = \omega R 
\ee
From which we read off the solution as
\be
\beal
R &= \omega^{-1}\cos \omega \lambda\\
\tau_- &=  \omega^{-1}\sin \omega \lambda\\
\tau_+ &=  \ell \text{arctan}\left [\frac{\sin \omega \lambda}
{\ell (\omega - 2 \bar{\sigma})} \right]
\eeal
\ee
Note that this solution has the size determined by $\omega^{-1}$, which is 
identical to the CDL bubble side, \eqref{RCDL}, found by analytically continuing
global de Sitter. In contrast however, the periodicity of this solution,
$\lambda \in [-\frac{\pi}{2\omega},\frac{\pi}{2\omega}]$, translates to 
$\Delta\tau_- =2/\omega$ and
$\Delta\tau_+ =2 \ell \text{arccot} [\ell (\omega - 2 \bar{\sigma})]$. However,
the de Sitter false vacuum in the static patch has a natural periodicity $2\pi \ell$.
Thus, the instanton computation will take into account the regularization of the conical
deficit integral implicitly, and can therefore be regarded as a check of this procedure.

Substituting this solution into \eqref{bounceaction}, noting that with $M_\pm=0$,
$f_+ = 1 - r^2/\ell^2$, $f_-=1$, and the entropy terms vanish, gives
\be
{\cal B}= \oint d\lambda \frac{R}{2G} (\dot{\tau}_+ - \dot{\tau}_-)
= \frac{\pi}{4G} \left ( \ell - \sqrt{\ell^2 - 1/\omega^2} \right)^2
= \frac{\pi \ell^2}{G} \frac{16 \bar{\sigma}^4 \ell^4}
{(1 + 4 \bar{\sigma}^2 \ell^2)^2}
\end{equation}
i.e.\ the CDL result \eqref{BCDL}. This result would not have been obtained without
the correct evaluation of the Einstein-Hilbert integral over the conical singularity.
Clearly, the result for a physical process should not depend on the choice of
coordinates for the analytic continuation. Here, we have analytically continued the
static time patch, whereas the canonical CDL computation analytically continues the
global time coordinate. These are different analytic continuations, but with the right
calculational method, we get the same result.

\subsubsection{Static bubbles}

Another example where we can find the tunnelling amplitude analytically is when the
instanton is independent of Euclidean time, $R(\lambda) = R_0$. In this case 
\eqref{propertime} implies that $f_\pm \dot{\tau}_\pm^2=1$. Then the individual 
components of the Israel equations
\be
\frac{f_+ \dot{\tau}_+ - f_-\dot{\tau}_-}{R}
= \sigma = \frac{f'_+ - 2 \ddot{R}}{2 f_+ \dot{\tau}_+} - 
\frac{f'_- - 2 \ddot{R}}{2 f_- \dot{\tau}_-}
\ee
can be seen to imply that the wall integral in \eqref{bounceaction} vanishes,
hence we are led to the remarkably simple result
\be
{\cal B}= {{\cal A}_+\over 4G}-{{\cal A}_-\over 4G} 
\label{staticB}
\ee
i.e.\ the probability of tunnelling from a seed to remnant black hole via a static 
instanton is simply the Boltzmann suppression due to the entropy difference:
\be
{\cal P} \sim e^{-\delta S/\hbar}
\ee
This turns out to be the key result of this work -- that black holes can give rise
to static instantons where the tunnelling rate is given by the the Boltzmann factor 
associated to the drop in entropy of the configuration.

\section{The Fate of the Higgs Vacuum}

The previous discussion of bubbles used the thin wall approximation. This was a
physically intuitive argument developed by Coleman to get analytic results quickly
in an era of minimal computing power. In reality, the Higgs potential does not have 
a form in which thin walls are even a reasonable approximation to a bubble. In this
section I review the thick wall bubbles relevant for Higgs vacuum decay.
I then discuss the ultimate fate of the Higgs vacuum, first comparing
the decay rate to other quantum processes with a black hole, then remarking
on the mutual constraint between vacuum metastability and cosmological
primordial black holes.

\subsection{From Thin to Thick: Higgs vacuum bubbles}

\begin{figure}[htb]
\begin{center}
\includegraphics[width=0.5\textwidth]{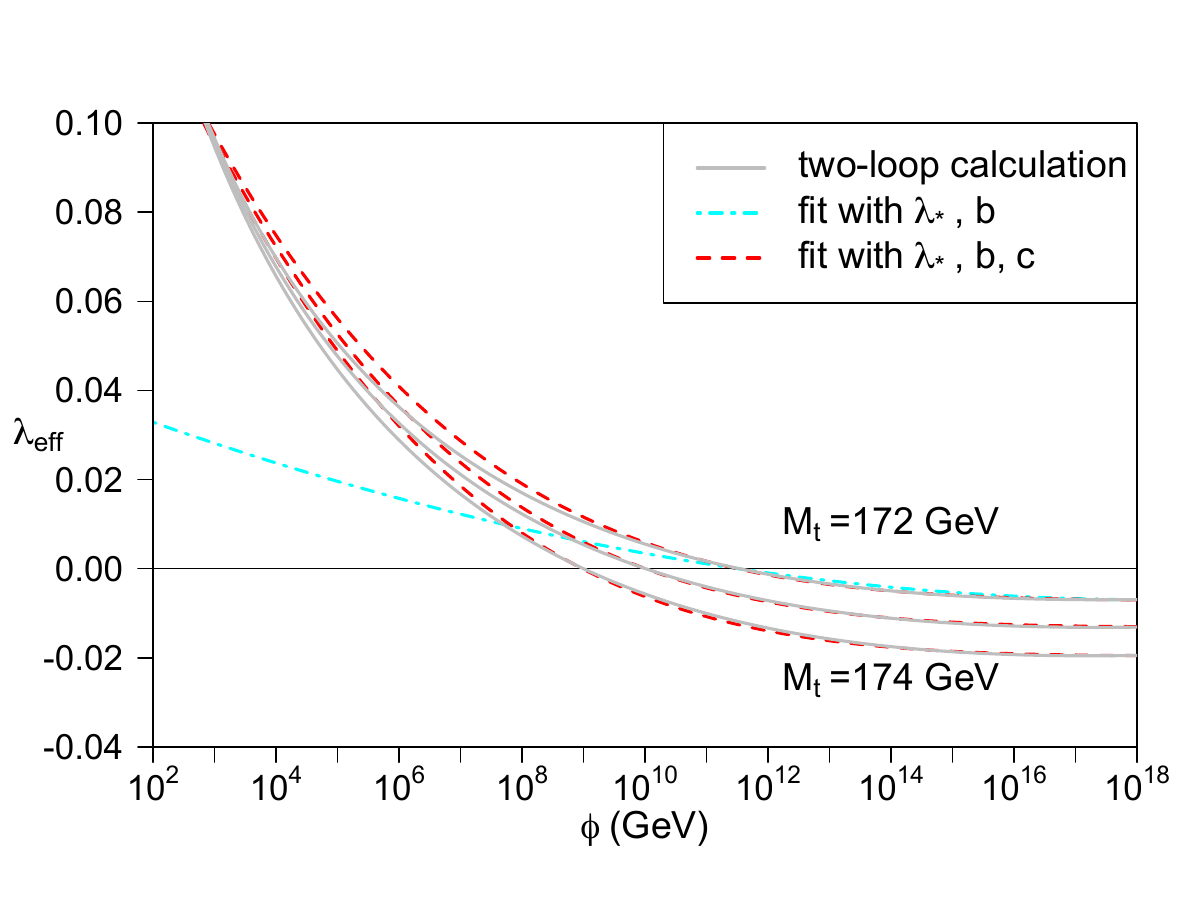}~~
\includegraphics[width=0.42\textwidth]{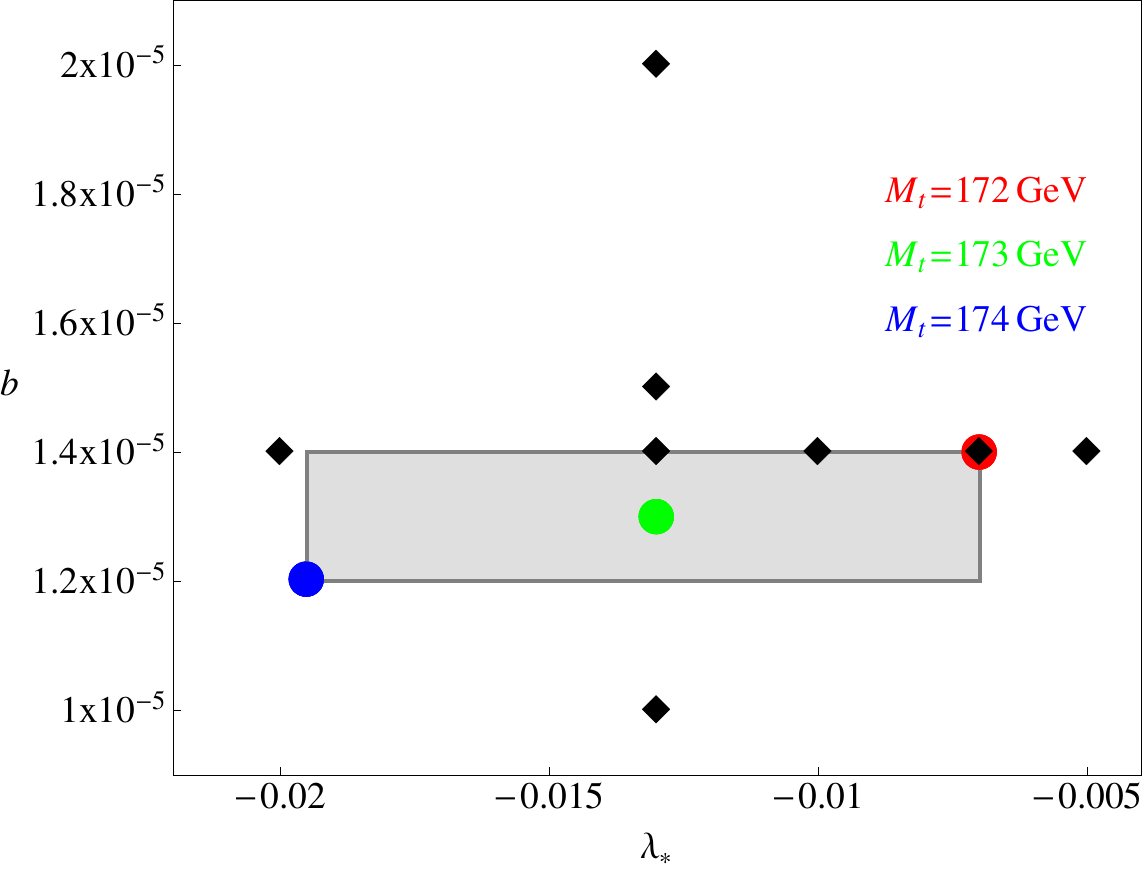}
\caption{Left:
The three parameter fit of the high-energy effective coupling used for vacuum 
decay results in \cite{Burda:2016mou}. All three parameters can be fixed by matching to
the SM calculation for a given Higgs and Top quark mass.
The plots show Higgs mass $M_{\rm H}=125{\rm GeV}$ and top quark masses
$172{\rm GeV}$ ($\lambda_*=-0.007$),  $173{\rm GeV}$ ($\lambda_*=-0.013$)
and $174{\rm GeV}$ ($\lambda_*=-0.00195$). A two parameter fit is shown for comparison.
Right: The values of $\lambda_*$ and $b$ panned over in the numerical integrations 
presented in the branching ratios in subsection \ref{sec:branch}.}
\label{fig:lambda}
\end{center}
\end{figure}

For shorthand, the Higgs potential is written as a function of the magnitude of
the Higgs, $\phi = |\Phi|$, and the potential is expressed in terms of
an effective quartic coupling constant $\lambda_{\rm eff}$,
\be
V(\phi)=\frac14\lambda_{\rm eff}(\phi)\phi^4
\ee
Radiative corrections have been computed 
\cite{Espinosa:2007qp,Ford:1992mv,Chetyrkin:2012rz,Bezrukov:2012sa,Degrassi:2012ry}, 
and in \cite{Burda:2015yfa,Burda:2016mou}, were approximated by the analytic 
three-parameter fit  
\be
\lambda_{\rm eff}(\phi)=\lambda_*+b\left(\ln{\phi\over M_p}\right)^2
+c\left(\ln{\phi\over M_p}\right)^4.
\label{higgspot}
\ee
which provides a good approximation to the exact results (which have to be computed
individually for given Top and Higgs masses) over a wide range of energy, and 
allows efficient numerical evaluation of the instanton equations of motion.
(See figure \ref{fig:lambda}.)

This potential does not admit thin wall solutions, so to link Higgs bubbles to the 
thin wall bubbles of the previous section, we add a
quantum gravity motivated correction to the scalar potential
\cite{Bergerhoff:1999jj,Branchina:2014rva,Eichhorn2015,
Greenwood:2008qp,Lalak:2014qua,Loebbert:2015eea}:
\be
V(\phi)=\frac14\lambda_{\rm eff}(\phi)\phi^4
+\frac16\lambda_6{\phi^6\over M_p^2}+\dots\label{hehiggs}
\ee
The effect of $\lambda_6$ is to add a definite minimum to the potential at large 
values of $\phi$ with sufficient barrier to give an approximate thin wall.
Using the observation that realistic bubbles have static instanton solutions, 
the Ansatz used in \cite{Burda:2016mou} was 
\begin{equation}
ds^2=f(r)e^{2\delta(r)}d\tau^2+{dr^2\over f(r)}+r^2(d\theta^2+\sin^2\theta d\varphi^2),
\end{equation}
where $f$ contains contributions from the black hole, the bubble, and any vacuum energy
\begin{equation}
f=1-{2G\mu(r)\over r}
\end{equation}
in the function $\mu(r)$. The redshift function $\delta(r)$ on the other hand
responds purely to the energy momentum of the bubble.
The equations of motion are
\bea
f\phi''+f'\phi'+{2\over r}f\phi'+\delta'f\phi' &=& V_\phi,\label{e1}\\
\mu'&=&4\pi r^2\left(\frac12f\phi^{\prime 2}+V\right),\label{e2}\\
\delta' &=&4\pi G r\phi^{\prime 2} \label{e3}
\eea
The equations for $\mu$ and $\phi$ can be decoupled from the $\delta$ equation,
which can be found from substituting the solution for $\phi$ in \eqref{e3}.
\begin{figure}[htb]
\begin{center}
\includegraphics[width=0.45\textwidth]{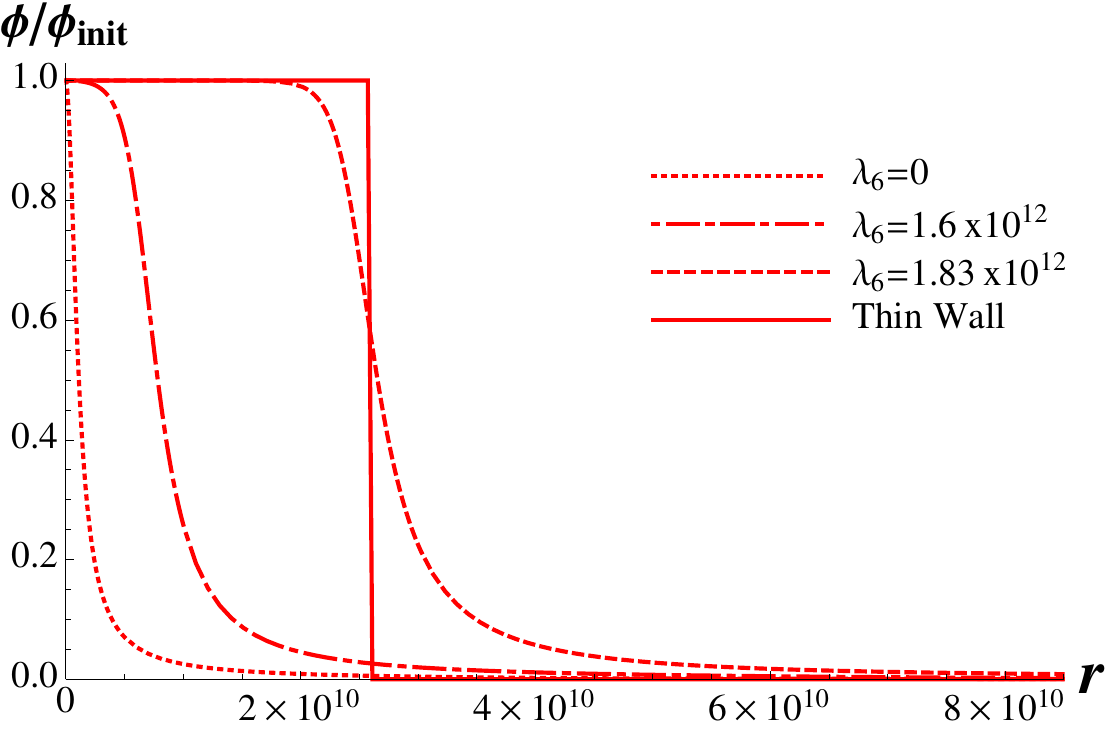}~~~
\includegraphics[width=0.45\textwidth]{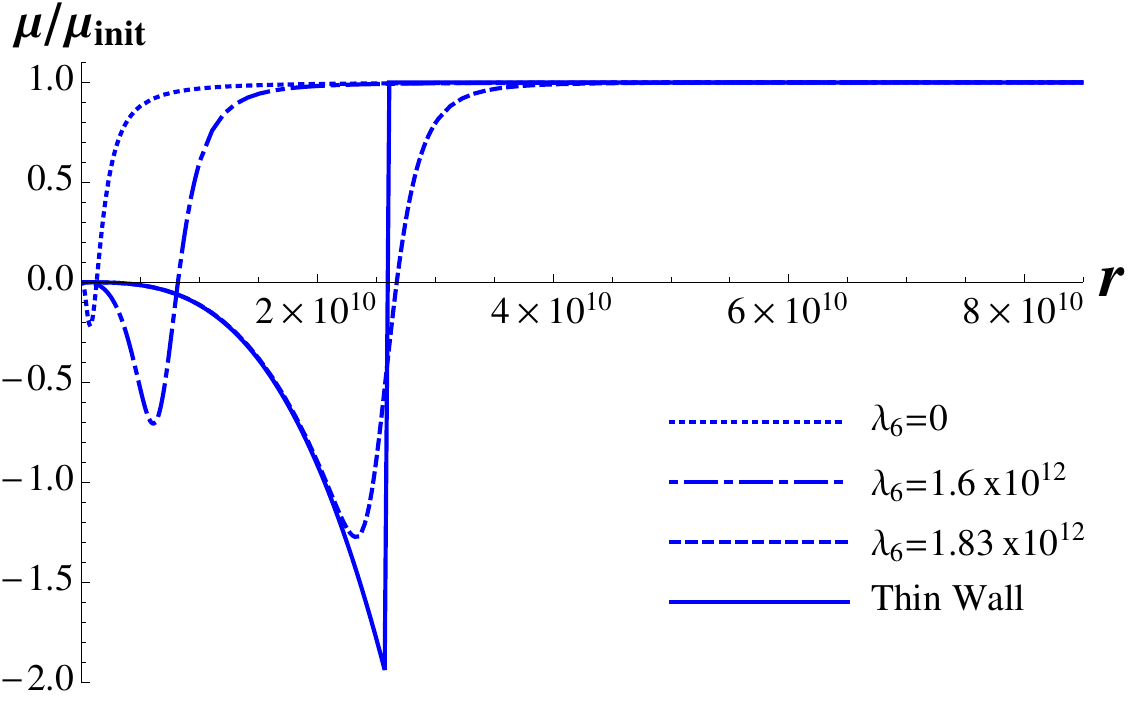}
\caption{
An illustration of the effect of switching on the $\lambda_6$ parameter. The
thin wall solution is shown as a solid line, with intermediate values of $\lambda_6$
shown. Note however the large values of the coupling that are necessary to 
produce anything approaching a thin wall.}
\label{fig:thickthin}
\end{center}
\end{figure}

Figure \ref{fig:thickthin} shows the solutions for the bounce, changing $\lambda_6$ 
to tune the bubble wall solutions from the thin wall of the previous section to the thick 
wall of the SM Higgs with $\lambda_6=0$. The full details of the process
are given in \cite{Burda:2016mou}.

Having found the solution, the exponent in the bounce amplitude is given by the entropy
shift between the seed and remnant black hole. In this case, considering the decay 
of the Higgs vacuum, the false vacuum energy is taken to be zero, therefore we are 
tunnelling to a negative vacuum energy. Writing $\mu(r_-)=\mu_-$ as the boundary
value of $\mu$ at the horizon of the remnant black hole, and $\mu_\infty=M_+$
as the asymptotic value of $\mu$ we have
\be
{\cal B}= \frac{{\cal A}_+}{4G}- \frac{{\cal A}_-}{4G} 
=  \frac{\mu^2_\infty - \mu_-^2}{2 M_p^2}
\ee
Thus, having found the numerical bubble solution, the bounce action is easily extracted 
in terms of the values of $\mu$ at each of the boundaries.

The main output of the numerical work in \cite{Burda:2016mou} is that the bubble
is diffuse, with the Higgs gradually varying from the SM value to large $\phi$ at the
remnant black hole horizon. Due to the weak coupling of gravity, this profile has
a relatively small effect on the mass function, so $(\mu_\infty-\mu_-)/M_p \ll 1$.
The vacuum decay rate has the form ${\cal A} e^{-{\cal B}}$, where ${\cal A}$ is a
pre-factor determined in field theory by the determinant of fluctuations around the saddle
point together with a factor of $\sqrt{{\cal B}/2\pi}$ for each translational zero mode 
\cite{callan1977}. Here, we have a single zero mode from the time-translation symmetry,
and estimate the determinant factor (the evaluation of the determinant is problematic
in Euclidean Quantum Gravity) by horizon timescale $(GM_+)^{-1}$. Putting this together, 
our estimate for the vacuum decay rate is:
\be
\Gamma_D \approx \left ( \frac{{\cal B}}{2\pi} \right) ^{\frac12} \left ( GM_+ \right )^{-1}
e^{-{\cal B}}
\ee
We now turn to the consequences of this result.

\subsection{A comparison with Hawking evaporation}
\label{sec:branch}

Having computed the vacuum decay rate in the presence of a black hole, we note
that, as is characteristic for a tunnelling process, it is exponentially suppressed. 
However, no matter how strongly suppressed, if the half life is significantly less 
than the age of the universe this is problematic! On the other hand, we also have
Hawking's result \cite{Hawking:1975vcx}, that black holes evaporate, and taking
Page's result for the evaporation rate \cite{Page} 
\be
\Gamma_H = \frac{\dot{M}}{M} \approx 3.6\times 10^{-4}(G^2 M_+^3)^{-1}
\ee
we find the branching ratio of the tunnelling rate to the evaporation rate is 
\be
\frac{\Gamma_D}{\Gamma_H}\approx 44 \, \frac{M_+^2}{M_p^2} \,
{\cal B}^{1/2}  e^{-{\cal B}}
\ee

Whether this branching ratio is larger than unity for semi-classical black holes
centres around the comparison between the power law pre-factors of $M_+/M_p$,
which must be large, and the size of the exponent ${\cal B}$ for those values.
Having already noted that the difference in the mass function is small for the thick 
Higgs bubbles, a rough estimate of the bounce action is 
${\cal B} \sim M_+ \delta M/M_p^2$, hence we can get a quick estimate
of the behaviour of this branching ratio 
\be
\frac{\Gamma_D}{\Gamma_H}\approx 44 \, 
\left ( \frac{M_+}{M_p} \right) ^{5/2} \, \sqrt{\frac{\delta M}{M_p}}
e^{-M_+ \delta M / M_p^2}
\ee
\begin{figure}[htb]
\begin{center}
\includegraphics[width=0.6\textwidth]{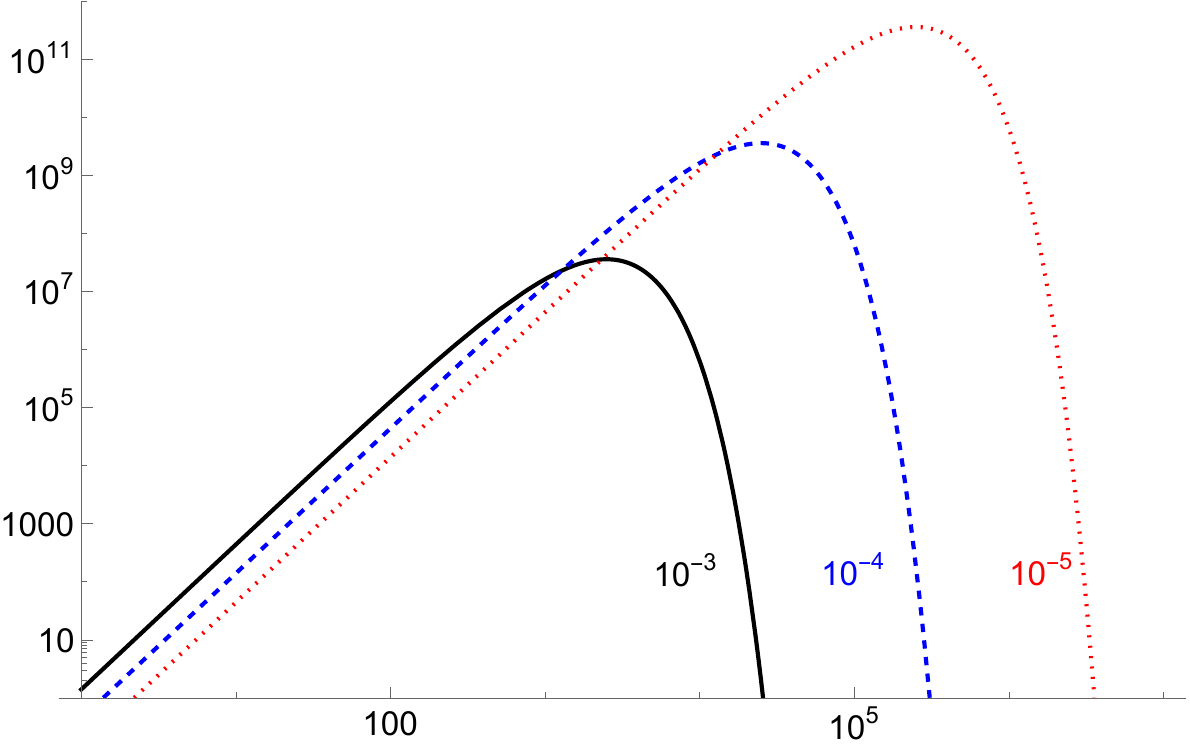}
\caption{
The branching ratio strongly depends on the difference in the mass function
generated by the diffuse bubble.}
\label{fig:branchapprox}
\end{center}
\end{figure}
Figure \ref{fig:branchapprox} shows a plot of this function, and how it depends on 
this mass difference, which itself in turn depends on the energy contained in the
diffuse Higgs bubble. While $M_+$ itself is large relative to the Planck mass 
(so that the spacetime curvature remains safely below the Planck scale), 
$\delta M$ will be much smaller, as it is dependent on the scale of the Higgs
relative to the Planck scale, integrated over the bubble.

As figure \ref{fig:branchapprox} shows eloquently, as $\delta M$ drops, 
the tunnelling rate is not just faster than evaporation, but becomes strongly 
dominant (by factors of $10^{10}$ or so) for seed masses around a gram.
The lifetime of a black hole of this mass is around $10^{-25}$s, hence if
the tunnelling process is strongly dominant, the decay will be essentially
instantaneous.

Having acquired some intuition on how important the vacuum decay process is
likely to be, we now show the actual results of \cite{Burda:2016mou} for the
vacuum decay branching ratios scanning through various fits to the SM
potential.
\begin{figure}[htb]
\begin{center}
\includegraphics[width=0.45\textwidth]{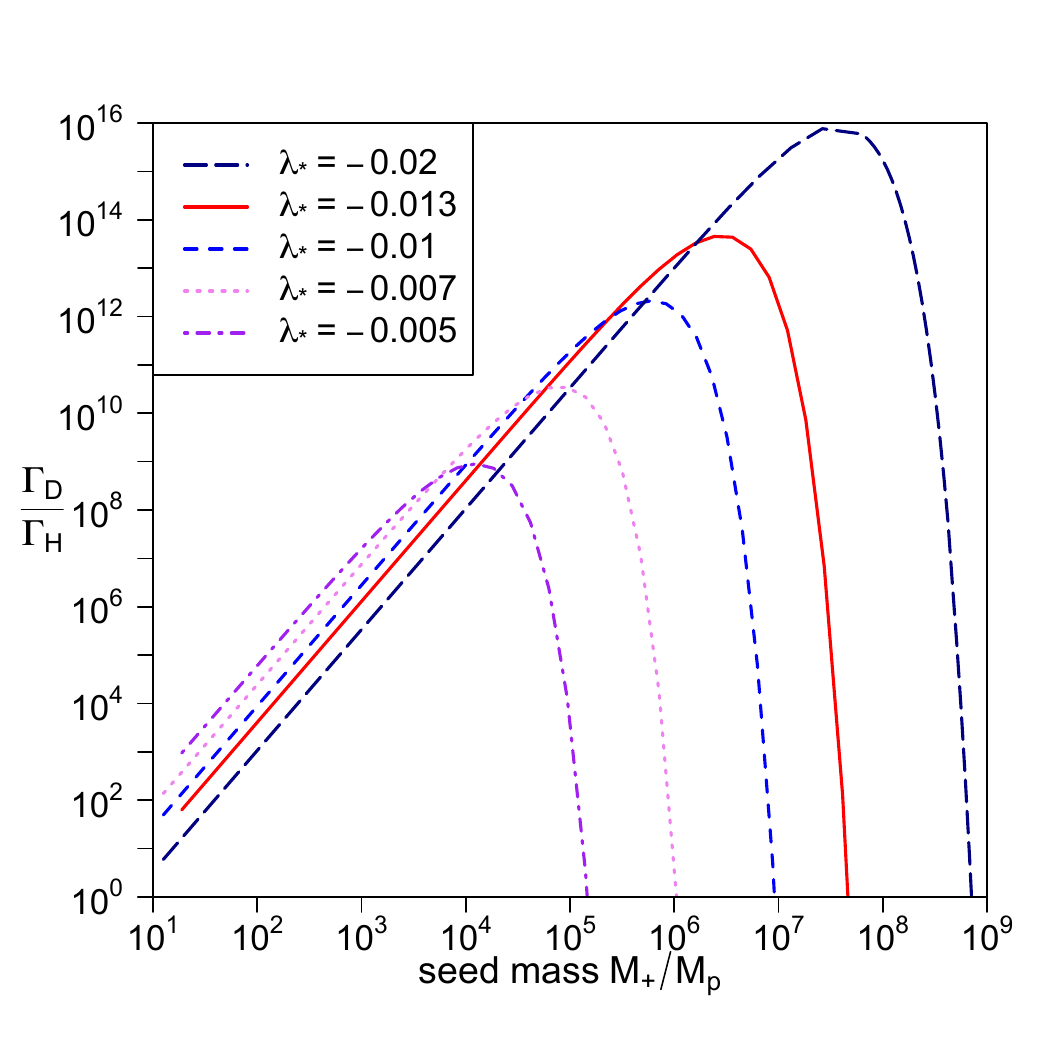}~~~
\includegraphics[width=0.45\textwidth]{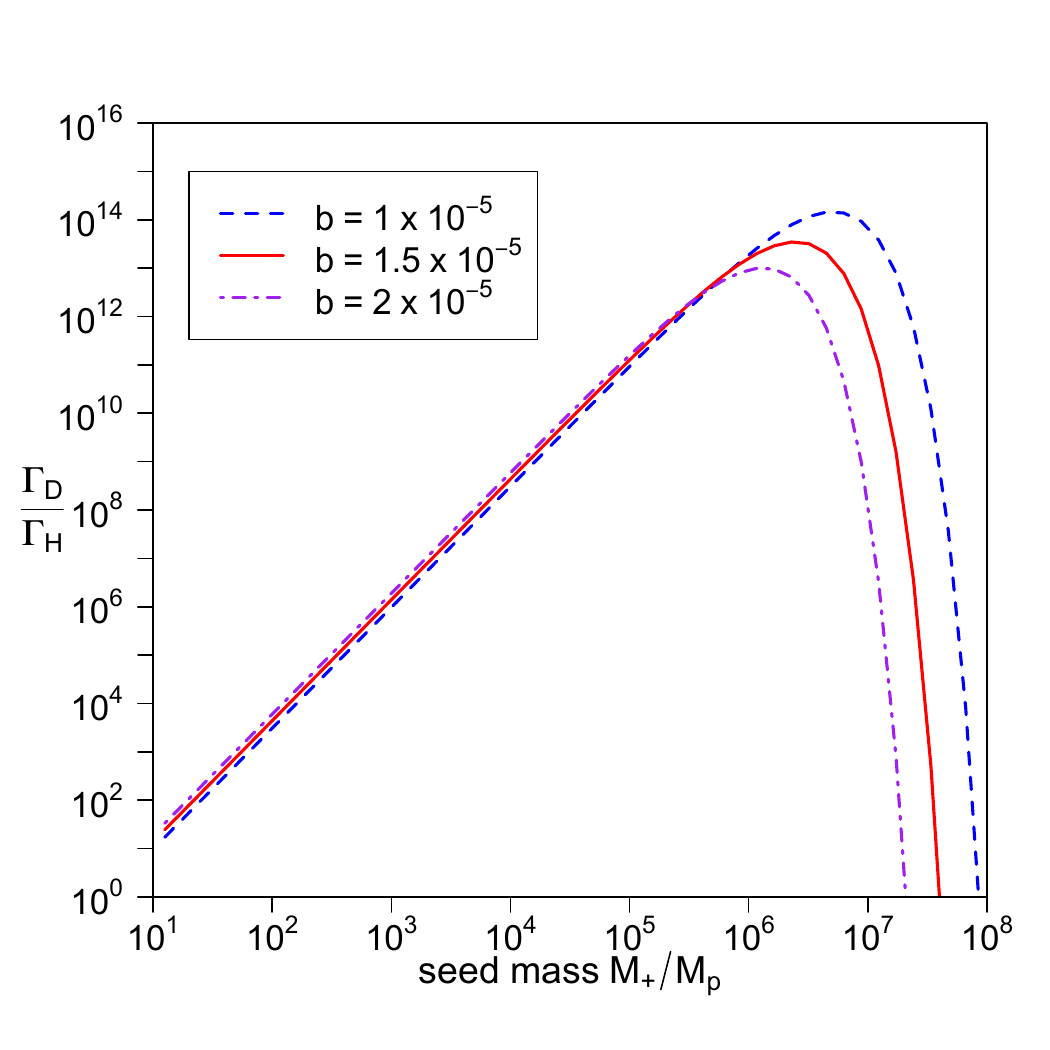}
\caption{
The branching ratio of vacuum decay to evaporation shown as a function of
seed mass for different values in the 3 parameter fit to the SM potential. 
On the Left, the impact of varying $\lambda_*$ with $b=1.4\times10^{-5}$. On
the Right, the impact of varying $b$ with $\lambda_*=-0.013$. (Note that
the value of the coupling at the SM scale then fixes $c$ in terms of $b$ and
$\lambda_*$.)}
\label{fig:SMrates}
\end{center}
\end{figure}
As figure \ref{fig:SMrates} shows, there is a wide range of black hole masses
for which vacuum decay is the dominant process. In terms of the three parameter
fit to the running Higgs coupling, the bare quartic coupling $\lambda_*$ has the 
biggest impact on the branching ratio, however it is clear that for a wide range of 
potentials, vacuum decay is inevitable if a black hole has small enough mass.
\begin{figure}[htb]
\begin{center}
\includegraphics[width=0.6\textwidth]{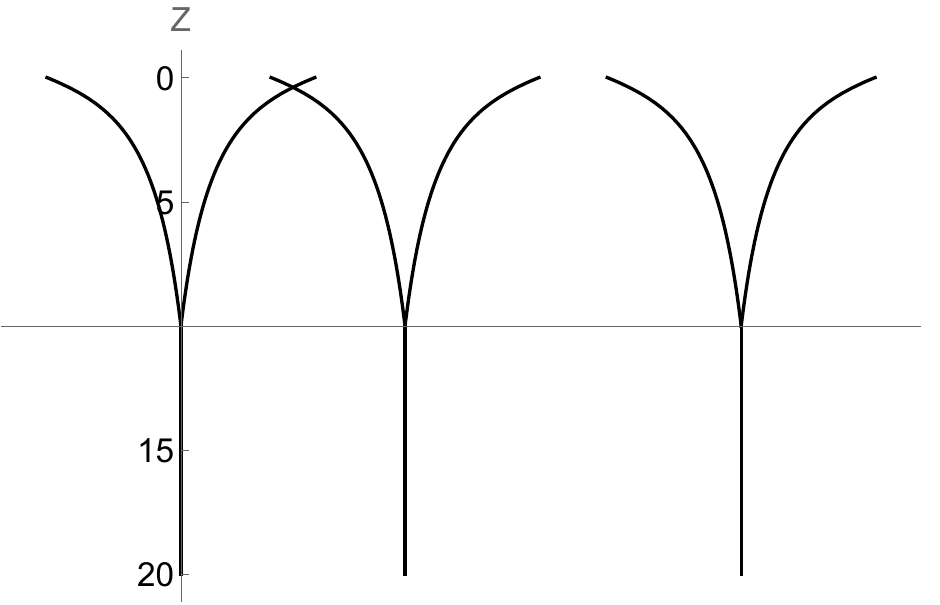}
\caption{
A sketch of the bubble percolation process for black holes reaching 
the critical mass range for vacuum decay at a redshift of 10.}
\label{fig:percolate}
\end{center}
\end{figure}

From \ref{fig:SMrates}, we see that in order to catalyse vacuum decay a black hole
has to be very light - of order a Kg or less! The only possibility for this type of black
hole is that it is a \emph{Primordial Black Hole} (PBH), i.e.\ one that is not formed by
gravitational mass of a star, but rather formed from strong perturbation
fluctuations in the very early universe \cite{Carr:1974nx}. 
Black holes formed at or below a mass of $\sim 4.5 \times 10^{14}$g would
have already lost sufficient mass by evaporation to lie within this range of
catalysis. Given that we are in the (presumed) metastable vacuum currently, 
we can deduce that there was no PBH with mass lighter than $\sim 4.5 \times 10^{14}$g
in our past light cone, however the question of whether black holes lighter
than this are incompatible with a metastable vacuum is a little more subtle, as
described in \cite{Dai:2019eei}, as the black hole not only has to seed decay, but the 
true vacuum would have to have percolated to encompass our current Hubble volume.
Figure \ref{fig:percolate} gives a sketch of how the bubbles would expand and percolate if 
formed at a redshift $Z =10$. The main outcome of the cosmological study is that
if PBH's are formed strongly peaked around a mass of $5 \times 10^{14}$ or less, then
this is incompatible with a metastable Higgs vacuum. 
(See also \cite{Cuspinera:2018woe,Mack:2018fny} for constraints from high 
energy particle collisions and extra dimensions.)

\section{Closing Thoughts}

In this presentation, I reviewed my work on vacuum decay catalysis by
black holes. In these examples, decay proceeds via bubble nucleation and is
described by a semi-classical Euclidean approach. One can have concerns
that this is not a well-defined approach for including gravity in quantum processes,
but then to be consistent one should never use the Euclidean method for computing
the temperature of a black hole. My opinion is that one either accepts \emph{all} 
the usual compromises for quantum processes with gravity, or seeks an alternative.
Given the success of the inflationary paradigm in explaining the perturbation spectrum
of the microwave background, this would be a rather extreme point of view!
Nonetheless, the result that having an impurity enhances and catalyses decay
is very much in keeping with the way first order phase transitions manifest in nature.

Other explorations of the impact of black holes on tunnelling processes include 
the Hawking Moss instanton \cite{Hawking:1981fz} -- an even more perplexing 
quantum process in which the entire universe up-tunnels to a higher cosmological 
constant before relaxing to a lower vacuum energy than the initial state. Here,
the black hole also enhances the tunnelling probability 
\cite{Gregory:2020cvy,Gregory:2020hia} while at the same time raising
interesting questions around constraints on allowed vacuum transitions.
Tunnelling in extended theories of gravity (e.g.\ 
\cite{Rajantie:2016hkj,Cuspinera:2019jwt,Gregory:2023kam}) both with
and without seeds has also been studied, with broadly similar outcomes.

The use of the regularisation of singular instantons raises questions around
analytic continuation and whether the thermal nature of the environment
has fully been taken into account (see e.g.\ \cite{Kohri:2017ybt,Hayashi:2020ocn,
Shkerin:2021rhy,Shkerin:2021zbf,Briaud:2022few,He:2022wwy}).
What these results generally indicate is that the Euclidean method is somehow dissatisfying 
as a description of quantum tunnelling, and that perhaps the time is ripe to
consider alternative approaches to describing tunnelling in both field theory and 
gravity (see, e.g.\ \cite{Braden:2018tky,Hayashi:2021kro,Feldbrugge:2022idb})
or indeed to testing tunnelling, and the implications of seeds, in the lab 
\cite{Fialko:2014xba,Braden:2017add,Billam:2018pvp,Zenesini:2023afv}.

Ultimately, whether or not the Standard Model is metastable, the challenge to critique
our set of tools and methodology in semi-classical gravity and field theory
gives a timely and compelling impetus to re-examine our description
of nonperturbative processes, and hopefully will lead to new and deeper
understanding.

%
%
\section*{Acknowledgement}
It is a pleasure to thank all my collaborators on this program of work:
Tom Billam, Philipp Burda, Leo Cuspinera, De Chang Dai, ShiQian Hu,
Katie Marshall, Florent Michel, Naritaka Oshita, Sam Patrick, Dejan Stojkovic,
Ben Withers, and a special thanks to my primary collaborator, Ian Moss.
This work has been supported by the STFC grants ST/P000371/, ST/J000426/1.
I also acknowledge the Aspen Center for Physics, supported by National Science 
Foundation grant PHY-2210452, and the Perimeter Institute for Theoretical Physics. 
Research at Perimeter Institute is supported by 
the Government of Canada through the Department of Innovation, 
Science and Economic Development Canada and by the Province of 
Ontario through the Ministry of Colleges and Universities.

%

%
%
%

\end{document}